\documentclass[a4paper,english,prl,twocolumn,showpacs]{revtex4}
\usepackage{ae} 
\usepackage[T1]{fontenc}
\usepackage[ansinew]{inputenc}
\usepackage{amsmath}
\usepackage{amssymb}
\usepackage{amsfonts}
\usepackage[dvips]{graphicx}
\usepackage{ifthen}
\usepackage{eurosym}
\usepackage{babel}

\newcommand{\Ps}{Ps$_{2}$}
\newcommand{\KA}[1]{\left< #1 \right>}
\newcommand{\ka}[1]{\langle #1\rangle}

\renewcommand{\d}{\text{d}}
\begin{document}

\title{Thermal dissociation of dipositronium: 
path integral Monte Carlo approach}
\author{Ilkka Kyl\"{a}np\"{a}\"{a} and Tapio T.~Rantala}
\affiliation{Tampere University of Technology, Department of Physics,
P.O. Box 692, FI-33101 Tampere, Finland}
\date{\today}

\begin{abstract}
Path integral Monte Carlo simulation of the dipositronium "molecule"
Ps$_2$ reveals its surprising thermal instability.  Although, 
the binding energy is $\sim 0.4$ eV, due to the strong
temperature dependence of its free energy Ps$_2$ dissociates, or
does not form, above $\sim 1000$ K, except for high densities
where a small fraction of molecules are in equilibrium with Ps atoms.
This prediction is consistent with the recently reported first
observation of stable Ps$_2$ molecules by
Cassidy \& Mills Jr., Nature {\bf 449}, 195 (07), and
Phys.Rev.Lett.~{\bf 100}, 013401 (08); at temperatures below $1000$ K.
The relatively sharp transition from molecular to atomic equilibrium,
that we find, remains to be experimentally verified.
To shed light on the origin of the large entropy factor in free energy 
we analyze the nature of interatomic interactions of these
strongly correlated quantum particles.
The conventional diatomic potential curve is given by the
van der Waals interaction at large distances, but due to the
correlations and high delocalization of constituent particles the 
concept of potential curve becomes ambiguous at short 
atomic distances.
\end{abstract}

\pacs{31.15.xk, 36.10.Dr, 31.15.ae}

\maketitle

Dipositronium or positronium molecule, Ps$_2$, is a four-body system
consisting of two electrons and two positrons. The dynamical stability
of dipositronium was established in 1947 by Hylleraas and Ore
\cite{Hylleraas47}. However, the molecule was not observed
experimentally until recently \cite{Cassidy_nature07}, even though a
lot of knowlegde had been provided by a number of theoretical studies,
see Refs.~\cite{Kinghorn93,Poshusta96,Bressanini97pra,Usukura98,
Usukura02,Schrader04,Bailey05pra} and references therein. In addition
to the fundamental issues of physics, \Ps{} is of interest also
in astrophysical applications and solid state physics
\cite{Bubin06pra,Emami-Razavi08pra}.

In laboratory conditions, \Ps{} formation has recently been
observed resulting from implantation of intense pulses of positrons
into porous silica films
\cite{Cassidy_nature07,Cassidy08prl}. 

The positronium molecule, with all the four particles of the same
mass, sets challenges to modeling, since quantum calculations are to
be performed fully non-adiabatically \cite{Kozlowski93}. This,
however, can be realized with quantum Monte Carlo (QMC) methods
\cite{Traynor91,Bressanini97,Kylanpaa07}. It should be pointed out
that also for other systems, approaches that are not restricted by the
Born--Oppenheimer or other adiabatic approximations are gaining more
attention
\cite{Kylanpaa07,Bhatia98,Taylor99,Korobov01,Gross01,Ohta03,Chakraborty08,Leeuwen08}.

Among the QMC methods the path integral Monte Carlo (PIMC) offers a
finite temperature approach together with a transparent tool to trace
the correlations between the particles involved.  Though
computationally challenging, with the carefully chosen approximations
PIMC is capable of treating low-dimensional systems, such as small
molecules or clusters accurately enough for good quantum statistics
for a finite temperature mixed state
\cite{Li87,Ceperley95,Pierce99,Kwon99,Knoll00,Cuervo06}.

In this study, using PIMC we evaluate the density matrix of the full
four-body quantum statistics in temperature dependent stationary
states.  Thus, the temperature dependent distributions of structures
and energetics of \Ps{} are established.  The main focus here is to find
the preferred configuration of the four-body system at each
temperature --- \Ps{} molecule or two Ps atoms.

According to the Feynman formulation of the statistical quantum
mechanics \cite{Fey72} the partition function for interacting
distinguishable particles is given by the trace of the density
matrix $\hat{\rho}(\beta) = e^{-\beta\hat{H}}$,
\begin{align}
Z
= \text{Tr}~\hat{\rho}(\beta)
= \int \d R_{0}\d R_{1} \ldots
\d R_{M-1} \prod_{i = 0}^{M-1}e^{-S(R_{i},R_{i+1};\tau)},\nonumber
\end{align}
where  $S$ is the action,
$\beta = 1/k_{\text{B}}T$, $\tau = \beta/M$ and $M$ is
called the Trotter number ($R_{M}=R_{0}$). In present simulations,
we use the pair approximation of the action and matrix squaring
for evaluation of the Coulomb interactions
 \cite{Storer68,Ceperley95}. Sampling of the paths in
the configuration space is carried out using the Metropolis algorithm
\cite{Metro53} with the bisection moves \cite{Chakravarty98}. The
Coulomb potential energy is obtained as an expectation value
from sampling and the
kinetic energy is calculated using the virial estimator \cite{Herman82}.

The error estimate for the PIMC scheme is commonly given in powers of
the imaginary time time-step $\tau$ \cite{Ceperley95}.  Therefore, in
order to determine comparable thermal effects on the system we have
carried out the simulations with similar sized time-steps regardless
of the temperature. This way the temperature dependent properties can
be compared avoiding temperature dependent systematic errors.  The
standard error of the mean (SEM) with two-sigma limits is used to
indicate the statistical uncertainty, where relevant.  The average of
the chosen time-step is $\ka{\tau}\approx 0.0146 E_\text{H}^{-1}$,
where $E_\text{H}$ denotes the atomic unit of energy, Hartree
($\approx 27.2$ eV).  The other atomic unit we use here is Bohr radius
for the length, $a_0$ ($\approx 0.529$ \AA).

The total energy of positronium "atom" Ps is $-0.25$ at $0$ K and the
binding energy of the molecule \Ps{} is $0.0160$ ($\approx 0.435$
eV)~\cite{Schrader04}.  We find these values as zero Kelvin
extrapolates from our simulations at low temperatures.  We point out
that with PIMC we evaluate energetics as statistical expectation
values from sampling with less accuracy than that from conventional
solutions of wavefunctions and the zero Kelvin data we obtain as
extrapolates, only.

\begin{center}
\begin{figure}[b]
\includegraphics[width=8cm,height=6cm]{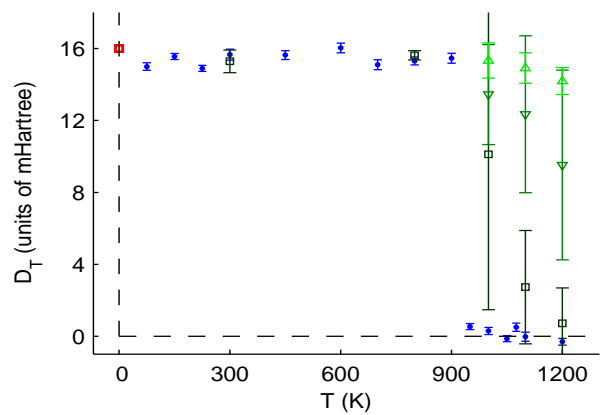}
\caption{\label{fig1}(Color online) Apparent temperature dependent
dissociation energy of dipositronium in units of mHartree: zero
Kelvin reference (red square) and finite temperature simulation
results at the low density limit (blue dots).
Data from higher Ps density simulations are also shown (green):
$0.50$ ($\square$), $14$ ($\bigtriangledown$) and 
 $100 \times 10^{24}$ m$^{-1}$ ($\bigtriangleup$).}
\end{figure}
\end{center}

In Fig.~\ref{fig1} we present the "apparent dissociation energy" of
\Ps{} at several different temperatures.  In each temperature this is
the negative total energy of the molecule with respect to two atoms as
$D_T = -[\langle E_{\rm tot}^{{\rm Ps}_2}\rangle_T -2 \langle E_{\rm
tot}^{{\rm Ps}}\rangle_T]$.  At $T \le 900$ K we find for the average
over shown temperatures $\bar{D}_T=0.0154(5)$, which is very close to
the dissociation energy at zero Kelvin, $D_0$.  However, at higher
temperatures the apparent dissociation energy vanishes,
because $\langle E_{\rm tot}^{{\rm Ps}_2}\rangle_T$ and
$2\langle E_{\rm tot}^{{\rm Ps}}\rangle_T$ become the same.  This is
because of molecular dissociation, or to be more exact, the two
atoms do not bind in our equilibrium state simulation at $T\ge 900$ K
and the predominant configuration is that of two separate positronium
atoms.

Simulations in a well-defined Ps density are time consuming,
and therefore, these kind of studies have been carried out
at the transition region around $1000$ K, only.
Using the periodic boundary conditions and the
cubic supercells of sizes from $(300 a_0)^3$ to $(50 a_0)^3$
with two Ps atoms we have simulated three densities from
$0.5$ to $100 \times 10^{24}$ m$^{-1}$, respectively.
We see that with increasing density the equilibrium shifts
to the molecular direction making the transition smoother
and raising it to higher temperatures compared to the
more sharp low density limit.

For completeness we should point out that in equilibrium
at any finite temperature the zero density limit consists
of Ps atoms, only.  Correspondingly, increasing density
will eventually smoothen the transition away.

In the recent experiment, cited above
\cite{Cassidy_nature07,Cassidy08prl}, formation of Ps$_2$
molecules was observed below $900$ K in about two orders of magnitude lower
densities than our lowest, above (Fig.~\ref{fig1}).  Formation
was not observed at higher temperatures, however, because the
Ps atoms desorbed from the confining porous silica
surface with the activation energy $k_BT \sim 0.074$ eV 
($\sim 850$ K).  Thus, our prediction of thermal dissociation
of Ps$_2$ above $900$ K in the experimentally achievable densities
remains to be verified in forthcoming experiments
in higher temperatures.

\begin{center}
\begin{figure}[t]
\includegraphics[width=8cm,height=6cm]{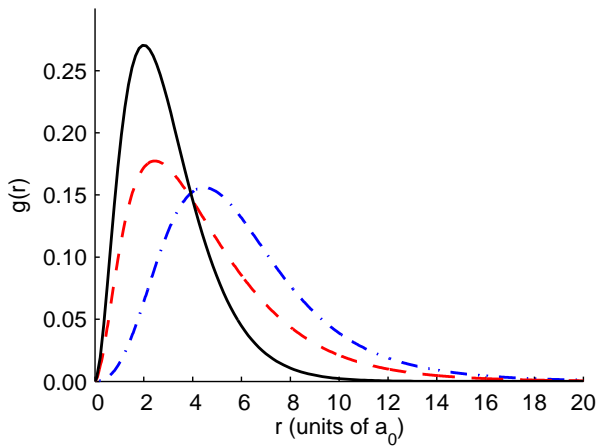}
\caption{\label{fig2}(Color online) Temperature averaged pair
correlation functions for different particle pairs ($T\le 900$ K):
e$^-$e$^-$ and e$^+$e$^+$ (dash dotted) and e$^-$e$^+$ (dashed). The
ground state ($T=300$ K) radial distribution of the free positronium
atom is given as a reference (solid line). The pair correlation
functions are averaged over temperatures below $900$ K.  The
distributions include the $r^2$ weight and normalization to one to allow
direct comparison to other published data, see Table \ref{table1}.}
\vskip -0.3cm
\end{figure}
\end{center}

\begin{table}[b]
\vskip -0.3cm
\caption{\label{table1}Simulated and reference data
\cite{Kinghorn93,Bressanini97,Usukura98,Bailey05pra,Bubin06pra} in
atomic units. Our data is given as averages from temperatures $T\le
900$ K. Apart from the energy the values are calculated using the
averaged pair correlation functions shown in Fig. \ref{fig2}.
Electrons are labeled 1 and 2, positrons are 3 and 4. Because of
symmetry $\ka{r_{12}}=\ka{r_{34}}$ and
$\ka{r_{13}}=\ka{r_{23}}=\ka{r_{14}}=\ka{r_{24}}$.  }
\begin{ruledtabular}
\begin{tabular}{lccccccc}
 & $\ka{E_{\text{tot}}}$ & $\ka{r_{12}}$ & $\ka{r_{13}}$ & $\ka{r_{12}^{-1}}$ & $\ka{r_{13}^{-1}}$ & $\ka{r_{12}^2}$ & $\ka{r_{13}^2}$\\
\hline
 Refs. & $-0.5160$ &  $6.033$ &  $4.487$ &  $0.221$ &  $0.368$ & $46.375$ & $29.113$ \\
 PIMC & $-0.5154(5)$ & $6.02$ & $4.48$ & $0.22$ & $0.37$ & $45.67$ & $28.78$\\
\end{tabular}
\end{ruledtabular}
\end{table}

\vskip -1.0cm  

Next, we compare our finite-temperature Ps$_2$ data to the published
zero Klevin results, discuss the details of Ps--Ps interaction, and finally,
conclude with the explanation of the higher temperature instability.

The conventional zero Kelvin like \Ps{} state of the system below
$900$ K is confirmed from the distributions in Fig.~\ref{fig2} and
related data in Table \ref{table1}.  The pair correlation functions
for like and opposite charged particles are essentially identical with
those reported elsewhere \cite{Usukura98} and the expectation values
of various powers of these distributions match with other published
reference data.  At higher temperatures, where $D_T \approx 0 $ K, the
corresponding distributions and data become that of the free Ps
atoms.

At $900$ K the thermal energy $k_B T = 0.0030 E_\text{H} \approx 0.08$
eV, only.  Therefore, the obvious question arises: Why the Ps$_2$
molecule with binding energy $0.44$ eV is unstable above $900$ K?  Is
there a temperature dependence hidden in the interactions? What does
the potential energy curve of this diatomic molecule look like?

It is the van der Waals interaction or sc.~dispersion forces,
that are expected to contribute to the potential curve
at larger atomic distances.
These arise from the "dynamic dipole--dipole correlations",
as usually quoted.   Now, within our approach we have a transparent
way to consider these interactions:
the dipoles and their relative orientations.
Thus, we monitor the dipole--dipole orientation correlation function
\begin{align}
  \label{eq:Orient}
  \KA{\frac{{\bf p}_I\cdot{\bf p}_J}{p_Ip_J}}
\end{align}
as a function of interatomic distance $R$,
where ${\bf p}_I$ and ${\bf p}_J$ are the two ${e^-e^+}$ dipoles.
This function assumes values from $-1$ to $0$, corresponding
orientations from perfectly opposite to fully random.

The concept of interatomic distance needs to be defined for evaluation.
We should note that at the "equlibrium distance" the
centers-of-mass (CM) of all four particles are superimposed on the
same location, as evaluated from their one particle distributions (or
wavefunctions).  However, the particles do have well-defined
(correlated) average distances, see Table \ref{table1}.
Thus, the definition is not trivial.

We can define the center-of-mass interatomic distance $R_{\rm CM}$
using the expectation value of the CM of one ${e^-e^+}$ pair and
that of the other pair.  An alternative (correlated) definition is the
expectation value of the separation of the two ${e^-e^+}$ dipoles,
$R_{\rm dd}$.  At large distances these two coincide, but
at the opposite limit, in \Ps{} molecule, the former becomes zero whereas the
latter remains at about 4 $a_0$.

Another problem is that
in an equilibrium simulation we are not able to choose or
fix the interatomic distance $R$ ($R_{\rm CM}$ or $R_{\rm dd}$).
Therefore, evaluation of $R$ dependent quantities presumes that
sampling in the chosen temperature includes the relevant $R$
with good enough statistics.  This kind of data hunting turns out to be
computationally challenging.

To overcome this, we have used a
"close-to-equilibrium" technique by starting from $800$ K distribution
and rising the temperature to $1000$ K, and then, applying the reverse
change in temperature to obtain another estimate.
In the former case we are able to follow the increase in $R$
from the molecular region to "dissociation", while the latter follows
"recombination".

\begin{center}
\begin{figure}[t]
\includegraphics[width=8cm,height=4.5cm]{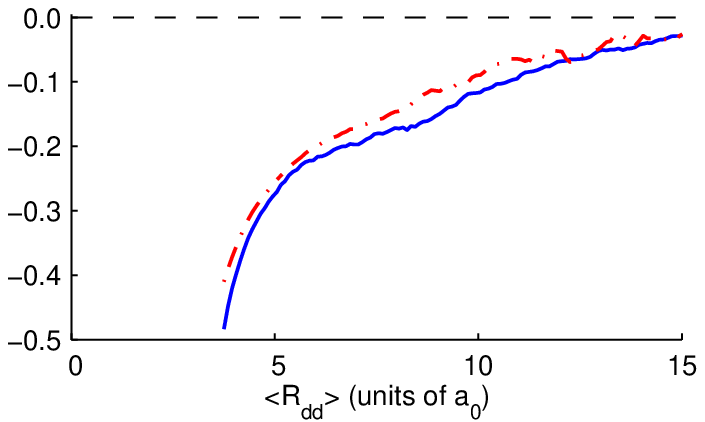}
\includegraphics[width=8cm,height=4.5cm]{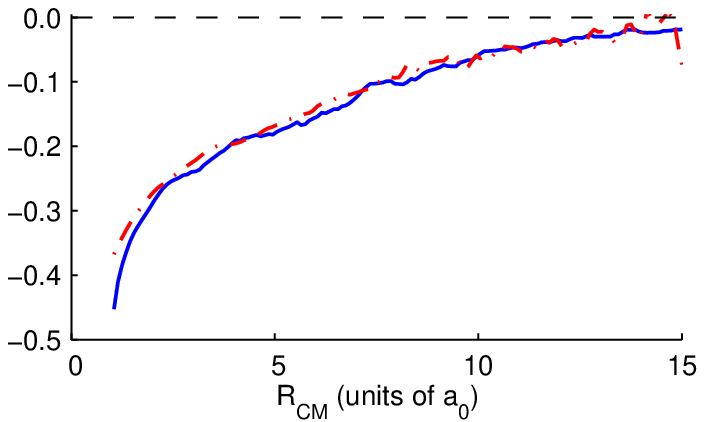}
\caption{\label{fig3}(Color online)
Dipole--dipole correlation functions,
Eq.~\eqref{eq:Orient}.
The upper (dash-dotted, red) and lower (solid, blue) curves correspond
to $1000$ K and $800$ K, respectively.
See the two definitions of the interatomic distances $R_{\rm dd}$
and $R_{\rm CM}$ in text.}
\vskip -0.3cm
\end{figure}
\end{center}

In Fig.~\ref{fig3} we show the estimates from these two temperatures
to the correlation function wrt.~the interatomic distances $R_{\rm
dd}$ and $R_{\rm CM}$.  We emphasize that these are estimates, only,
because at different temperatures the equilibrium sampling regions of
$R$ are very different.  However, we see that the difference between
these two estimates is very small and the equilibrium simulation
correlation function between these two is easily conceived.  Thus, we
conclude that the dipole--dipole correlation is not temperature
dependent.

\begin{center}
\begin{figure}[b]
\includegraphics[width=8cm,height=4.5cm]{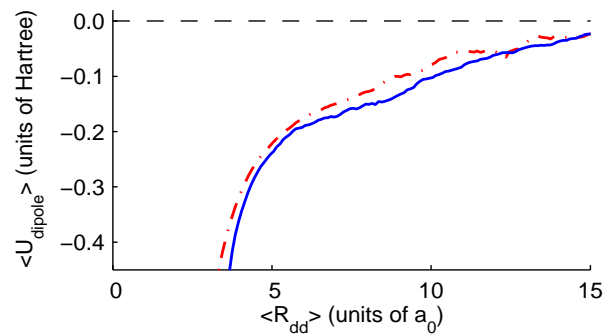}
\caption{\label{fig4}(Color online) Dipole--dipole interaction energy
with the same notations as in Fig.~\ref{fig3}.  The upper (dash-dotted,
red) and lower (solid, blue) curves correspond to $1000$ K and $800$ K,
respectively.}
\vskip -0.3cm
\end{figure}
\end{center}

\vskip -1.8cm

Using the same "close-to-equilibrium" technique we evaluate the van
der Waals interaction energy, next.  This is shown in Fig. \ref{fig4}.
There too, the true equilibrium curve can be estimated as the average
of the two shown ones.  Simple fit reveals that the large distance limit
($R_{\rm dd} > 12 a_0$) shows the asymptotic $R^{-\alpha}$ behavior
($\alpha$ roughly $6$) as expected.

Sampling all the energy contributions with the same
"close-to-equilibrium" technique allows us to evaluate the total
energy or the diatomic potential energy curve as a function of
interatomic distance $E_{\rm tot}^{{\rm Ps}_2}(R)$, where $R=$ $R_{\rm
dd}$ or $R_{\rm CM}$.  It shows the same temperature independent
behavior, though the statistics is not good enough to allow showing the curve,
here.  As expected, we find that the true dissociation
energy is not temperature dependent, as is the "apparent dissociation
energy" $D_T$ shown in Fig.~\ref{fig1}.

Now, the "thermal dissociation" can be explained by the strong
temperature dependence of the \Ps{} free energy.  With the rising
temperature the free energy of the two atoms decreases below that of
the molecule, leading to transition from the molecular dominance to
the atomic one.  This is not a surprise, but the usual behavior of the
conventional molecules.  From our simulations we find, however, the
following surprising features: (i) the low temperature, where the
transition takes place, (ii) sharpness of the transition and (ii)
almost negligible density dependence at the experimentally relevant
densities.

The transition temperature is usually estimated by matching
the thermal energy $k_B T$ with the dissociation energy.
This is where the entropic contribution in free energy $-TS$
becomes comparable with the dissociation energy.
In the present case, this gives about $5000$ K.
Conventionally, the transition is smooth following from
the equilibrium between molecular dissociation and formation,
where the former depends on the temperature, and the latter,
on the density, the density being the main factor in the entropy.

The Ps$_2$ molecule lacking in the heavy nuclei is peculiar.
All of its constituents are strongly delocalized,
barely fitting into the binding regime of the
molecular potential curve.  This is what they do below
$900$ K in experimentally relevant densities, but not above
$1000$ K.  This is a consequence from the exceptionally
large entropy factor originating from the strong quantum
delocalization more than the density.

In summary, with path integral Monte Carlo simulations of the
dipositronium "molecule" Ps$_2$ we have found and explained its
surprising thermal instability.  Due to the strong
temperature dependence of the free energy of the considered four particle
system the molecular form is less stable than two positronium atoms above about
$900$ K, though the molecular dissociation energy is $\sim 0.4$ eV.
The transition in equilibrium from molecules to atoms
is sharp in temperature and only weekly density dependent.
This can be understood by the large entropy factor originating
from strong delocalization of all of the molecular constituents.
Our prediction remains to be experimentally verified.

We thank David Ceperley for his attention and interest in our work.
For financial support we thank the Academy of Finland, and for
computational resources the facilities of Finnish IT Center for
Science (CSC) and Material Sciences National Grid Infrastructure
(M-grid, akaatti).

\bibliography{viitteet}

\end{document}